\documentclass[12pt, a4paper]{article}

\usepackage[pdftex]{graphicx}
\usepackage{bm,afterpage,amsmath,placeins,flafter,authblk,adjustbox,lscape,caption}
\usepackage[english]{babel}
\usepackage{setspace}
\usepackage{booktabs}
\usepackage{changes}
\usepackage[top=1.3in,left=1.1in,right=1.1in,bottom=1.3in]{geometry}
\usepackage{color}

\graphicspath{{figures/}}

\date{}

\begin{document}	
	\title{Sensor Analytics in Basketball}
	
	\author[1]{Rodolfo Metulini}
	\author[1]{Marica Manisera}
	\author[1]{Paola Zuccolotto}
	\affil[1]{\small Department of Economics and Management, University of Brescia}
	
	\maketitle

\textbf{Please cite as:} \textit{Metulini, R., Manisera, M., Zuccolotto, P. (2017), Sensor Analytics in Basketball. Proceedings of the 6th International Conference on Mathematics in Sport. ISBN 978-88-6938-058-7}

\begin{abstract}
 A new approach in team sports analysis consists in studying positioning and movements of players during the game in relation to team performance. State of the art tracking systems produce spatio-temporal traces of players that have facilitated a variety of research aimed to extract insights from  trajectories. Several methods borrowed from machine learning, network and complex systems, geographic information system, computer vision and statistics have been proposed. After having reviewed the state of the art in those niches of literature aiming to extract useful information to analysts and experts in terms of relation between players' trajectories and team performance, this paper presents preliminary results from analysing trajectories data and sheds light on potential future research in this field of study. In particular, using convex hulls, we find interesting regularities in players' movement patterns.
\end{abstract}


\section{Introduction}
Studying the interaction between players in the court, in relation to team performance, is one of the most important issues in sport science. Coaches and experts want to explain why, when and how specific movement is expressed because of tactical behaviour. Analysts want to study how cooperative movement patterns react to a variety of factors. To answer these questions, sports analysts borrowed methods from many disciplines, such as machine learning, network and complex systems, geographic information systems, computational geometry, computer vision and statistics. In recent years, the advent of information technology systems made it possible to collect a large amount of different types of big data, which are, basically, of two kinds. On the one hand, play-by-play data report a sequence of significant events, such as passes and shots, that occur during a match. A wide range of academic studies uses these data, with the aim of analysing team's performance: events collected in the play-by-play are used to identify the drivers that affect the probability to win a match, both from a data-mining perspective (\cite{Carpita et al. (2013),Carpita et al. (2015)}), and adopting social network analysis (\cite{Wasserman and Katherine  (1994), Passos et al. (2011), Cintia et al. (2015)}). On the other hand, object trajectories capture the movement of players using optical- or device-tracking and processing systems. These systems are based on Global Positioning Systems (GPS). The trajectory of a single player depends on the trajectories of all other players in the court, both team-mates and opponents, and on a large amount of external factors. A candidate method to approach with this complexity consists in segmenting a match into phases, to facilitate the retrieval of significant moments of the game (\cite{perin2013soccerstories}). Furthermore, a promising niche of literature called \textit{ecological dynamics} expresses players in the court as agents who face with external factors (\cite{travassos2013performance,Passos et al. (2011)}). Visualization tools are required in order to communicate the information extracted from trajectories. The growing interest in applying novel visual tools to a range of sports contexts has been highlighted by Basole and Saupe \cite{basole2016sports}. Visualization tools are used on leading outlets such as the \textit{New York Times} (\cite{aisch2016curry,goldsberry2013map}), and on academic journals (\cite{perin2013soccerstories,sacha2014feature,polk2014tennivis}).

\vline

We start by presenting the state of the art in those niches of sport science literature where the use of play-by-play and trajectories data is helpful to the aim of extracting information for analysts and experts (Section \ref{statart}). We position among these fields of literature with the final aim to study players' trajectories in basketball, in particular to visualize and characterize the movements of players around the court to find relevant types of movement patterns that could affect the team performance. To this aim, in Section \ref{datamet} we first present data and methods, while in Section \ref{question} we discuss in detail our research questions. We then present empirical analysis in Section \ref{results}. In Section \ref{concl} we conclude and discuss future research developments.

\section{State of the art} \label{statart}

\vline

\subsection{Data-driven approach}

\vline

Data-driven science, an interdisciplinary field about scientific methods, processes and systems, aims to extract insights from data in various forms. It employs techniques and theories drawn from many fields within the broad areas of mathematics, statistics, information science, and computer science, in particular from the sub-domains of machine learning, classification and data-mining. Data-driven approach benefits from the availability of the play-by-play and academic research is proliferating. Carpita et al. \cite{Carpita et al. (2013),Carpita et al. (2015)} used cluster analysis and principal component analysis in order to identify the drivers that affect the probability to win a football match. Social network analysis has also been used to capture the interactions between players. Wasserman and Faust \cite{Wasserman and Katherine (1994)} mainly focus on passing networks and transition networks. A passing network is a graph where each player is modelled as a vertex and successful passes to the player represent links among vertices. Transition networks can be constructed directly from event logs, and correspond to a passing network where play-by-play data are attached. Passos et al. \cite{Passos et al. (2011)} used centrality measures with the aim to identify central (or key) players, and to estimate the interaction and the cooperation between team members in water polo. In soccer, Cintia et al. \cite{Cintia et al. (2015)} observed players' behaviour on the pitch. They predict the outcome of a long-running tournament such as Italian major league using simple network measures. Moreover, Cintia et al. \cite{cintia2015harsh} proposed and computed a pass-based performance indicator that strongly correlates with the success of the team.

\vline

\subsection{Synchronyzed movements analysis}

\vline
	
The trajectory of a single player depends on a large amount of factors and on the trajectories of all other players in the court, both team-mates and opponents. Because of these interdependencies, in every single moment, a player action causes a reaction.  A promising niche of sport science literature, borrowing from the concept of physical psychology, Tuvey and Shaw \cite{turvey1995toward} expresses players in the court as agents who face with external factors (\cite{travassos2013performance,Passos et al. (2011)}). In addition, typically, players' movements are determined by their role in the game. Predefined plays are used in many team sports to achieve some specific objective; moreover, team-mates who are familiar with each other's playing style may develop ad-hoc productive interactions that are used repeatedly. On the one hand, experts want to explain why, when and how specific movement behaviour is expressed because of tactical behaviour. Brillinger \cite{brillinger2007potential} addressed the question of how to analytically describe the spatio-temporal movement of particular sequences of passes (i.e. the last 25 passes before a score). On the other hand, analysts want to explain and observe cooperative movement patterns in reaction to a variety of factors, such as coach advices and the corresponding team reactions, characteristics of the stadium (capacity, open/closed roof) and historical and current weather records (high/low temperatures, air humidity). A useful method to approach with complexity in team sport analysis consists in segmenting a match into phases, as it facilitates the retrieval of significant moments of the game. Perin et al.  \cite{perin2013soccerstories} visually segmented a footbal match into different phases while Metulini et al.  \cite{metulini2017cluster} segmented a basketball game into phases using a cluster analysis. A key factor in relation to teams' performance is how players control space. Many works are devoted to analyse how the space is occupied by players - when attacking and when defending - or in crucial moments of the match. Examples can be found in football (\cite{couceiro2014dynamical,moura2012quantitative}) or in futsal (\cite{fonseca2012spatial,travassos2012spatiotemporal}).
	 
\vline 
	 
\subsection{Visualization tools}

\vline

In order to communicate the information extracted from the spatio-temporal data, visualization tools are required. Basole and Saupe  \cite{basole2016sports} highlight the growing interest in applying novel visual tools to a range of sports contexts. Analysts from leading outlets such as the \textit{New York Times} use visualization to tell basketball and football stories (\cite{aisch2016curry,goldsberry2013map}). The increasing popularity of sports data visualization is also reflected in a greater academic interest. Perin et al. \cite{perin2013soccerstories} developed a system for visual exploration of phases in football, Sacha \cite{sacha2014feature} present a visual analysis system for interactive recognition of football patterns and situations. Notable works include data visualization in ice hockey (\cite{pileggi2012snapshot}) and tennis (\cite{polk2014tennivis}). In basketball, Losada et al. \cite{losada2016bkviz} developed `BKViz", a visual analytics system to reveal how players perform together and as individuals.  For visualizing aggregated information the most common approach is to use heat maps, simple and intuitive tools that can be used to visualize various types of data. Typical examples in the literature show the spread and range of a shooter (\cite{goldsberry2012courtvision}) or count how many times a player lies in specific court zones. More recently, dynamic approaches have been proposed to visualize aggregated information displaying the time dimension: Theron et al.  \cite{theron2010visual} employed tools for the analysis of players' movements. Metulini \cite{metulini2016motion} proposed the use of motion charts for visualizing movements of basketball players' in the court. There are several softwares providing the possibility to reproduce motion charts, more or less intuitive, open source or requiring a license (\textit{Gapminder world}, \textit{Google docs gadget}, \textit{Trend compass}, and JMP from SAS institute). In addition, motion charts can be created through web programming languages using \textit{Google application programming interface}, \textit{Google} API, \textit{Flash} or \textit{HTML5}. Applications of motion charts in other academic fields cover the aspects of students learning processes and linguistic changes (\cite{hilpert2011dynamic}), insurance (\cite{heinz2014pratical}) and development economics (\cite{saka2015inequality}), medicine (\cite{santori2014liver}) and hydrology (\cite{bolt2015water}.
	 
\section{Data \& Methods} \label{datamet}

\vline

\subsection{Global Positioning Systems (GPS)}

\vline

Object trajectories capture the movement of players (with or without data about the ball). Players' trajectories are retrieved using optical- or device-tracking and processing systems. Optical tracking systems use fixed cameras to collect the player movement, and the images are then processed to compute the trajectories (\cite{bradley2007reliability}). There are several commercial vendors who supply tracking services to professional sport teams and leagues (\cite{tracab2015,impire2015}). Tracking systems rely on devices that infer their location, and are attached to the players' clothing or embedded in the ball or puck. These systems are based on Global Positioning Systems (GPS) (\cite{catapult2015}). The resulting dataset is dense, because GPS collects data at very close instants. The adoption of this technology and the availability to researchers of the resulting data depends on various factors, particularly commercial and technical, such as, for example, the costs of installation and maintenance and the legislation adopted by the sport associations. This data acquisition may be partially restricted in some diffused team sports (as it was for example in soccer until 2015) while allowed for others.

\vline

\subsection{Play-by-play}

\vline

Play-by-play is a sequence of significant events that occur during a match. Events can be broadly categorised as player events such as passes and shots; and technical events, for example fouls, time-outs, and start/end of period.
Event logs are qualitatively different from the player trajectories in that they are not dense, as samples are only captured when an event occurs; however they can be semantically richer as they include details like the type of event and the players involved. Typically, in basketball, the play-by-play consists of a collection of about five hundreds events per game. The collection includes events such as made shots, missed shots, rebounds, fouls, start/end of the period, etc.. .

 \section{Research questions} \label{question}
 
The overall objective of our research is to visualize and characterize the movement of basketball players around the court by finding relevant types of movement patterns that could affect the team performance.

Going into detail, the first specific objective is to find and demonstrate the usefulness of a visual tool approach in order to extract preliminary insights from trajectories. In this respect, we aim to visualize the synchronized movement of players and to characterize the spatial pattern of them around the court in order to supply experts and analysts with useful tool in addition to traditional statistics, and to corroborate the interpretation of evidence from other methods of analysis. Some preliminary results in this respect have been found in Metulini \cite{metulini2016motion}, which suggested the use of motion charts to visualize the movements of players around the court and found, using sensor data from one game, differences in spacing structure among offensive and defensive plays. However, developments need to be carried on: analysing both team-mate and opponents trajectories together with the ball is fundamental in order to study how the movement of the players of one team reacts to the movement of the opposing team. 

Another aspect of research lies on segmenting the match into phases. Specifically, our idea is to find, through a cluster analysis, a number of groups each identifying a specific spatial pattern, in order to: i) characterize the synchronized movement of players around the court, ii) find any regularities and synchronizations in players' trajectories, by decomposing the game into homogeneous phases in terms of spatial relations. Some related exploratory analysis using one game data have been carried on by Metulini et al. \cite{metulini2017cluster}. We plan to extend the analysis to multiple matches. Moreover, we aim to match play-by-play data with trajectories, to extract insights on the relations between particular spatial pattern and the team performance. The effect of the ball's position in determining players' movement also deserves to be studied.

\section{Preliminary results} \label{results}

GPS trajectories data used in empirical analysis refers to a friendly match played on March 22th, 2016 by a team based in the city of Pavia (Italy). This team played the 2015-2016 season in the C-gold league, the fourth league in Italy. Totally, six players took part to the friendly match. All those players worn a microchip in their clothings. The microchip collected the position in both the $x$-axis and the $y$-axis. The positioning of the players has been detected at a resolution of  milliseconds. Considering all the six players recovered, the system recorded a total of $133,662$ space-time observations ordered in time. On average, the system collects positions about $37$ times every second. Considering all the six players, the position of each single player is collected, on average, every $162$ milliseconds. The dataset is structured such that \textit{tagid} variable uniquely identifies the player, \textit{timestamp\_ms} variable reports the exact millisecond in which the player position has been observed, \textit{klm\_x} and \textit{klm\_y}  represent the $x$-axis (length), and the $y$-axis (width) coordinates, filtered with a Kalman approach. The Kalman filtering is an algorithm that uses a series of measurements observed over time, in order to produce more precise estimates than than those based on a single measurement alone. 

We first use motion charts to visualize the synchronized spatio-temporal movements of players around the court. In our application, \texttt{gvisMotionChart} in $R$ is used (\cite{gesmann2013package}) because it outperforms alternatives in terms of open source, friendliness, and because it allows to import data.  A video tutorial showing players' trajectories using motion charts can be found at:

\texttt{http://bodai.unibs.it/BDSports/Ricerca2\%20-\%20DataInn.htm}.

To the best of our knowledge, motion charts have never been applied to basketball. Motion charts applied to our data show differences in the spacing structure of players among offensive and defensive plays. We defined whether each time instant corresponds to an offensive or a defensive play looking to the average coordinate of the five players in the court. The evidence is confirmed by the statistics describing the convex hull areas and the average distances, which are reported in Table \ref{tab:averages}.  Results clearly highlight larger average distances among players in offensive plays, as well as larger convex hull areas.
\begin{table}[htbp]
	\centering
	\caption{Average distances among players (in meters) and convex hulls areas (in squared meters) for the full match, for defensive and offensive plays.}
	\begin{tabular}{l|rrrr}	
		& \multicolumn{2}{c}{Average distances}              & \multicolumn{2}{c}{Convex hull area}  \\
		& attack & defence &       attack & defence \\
		\hline
		Min   & 2.296 & 0.400        & 1.000 & 1.000 \\
		1st Qu. & 6.372 & 4.309        & 30.000 & 14.000 \\
		Median & 7.235 & 5.086      & 41.000 & 20.500 \\
		Mean  & 7.250 & 5.680      & 42.590 & 28.550 \\
		3rd Qu. & 8.132 & 6.523        & 53.000 & 33.500 \\
		Max   & 13.947 & 14.260       & 138.500 & 180.000 \\
	\end{tabular}%
	\label{tab:averages}%
\end{table}%

\begin{figure}[!htb]
	\includegraphics[width=0.24\textwidth]{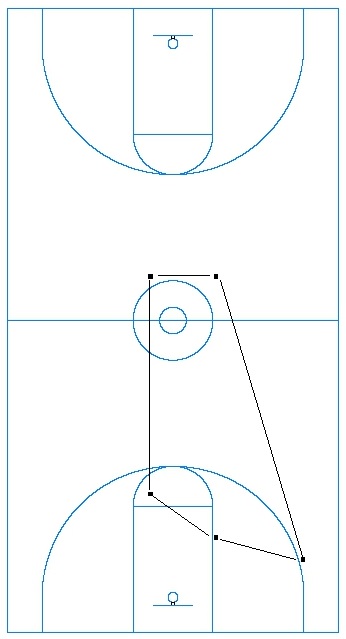}	
	\includegraphics[width=0.242\textwidth]{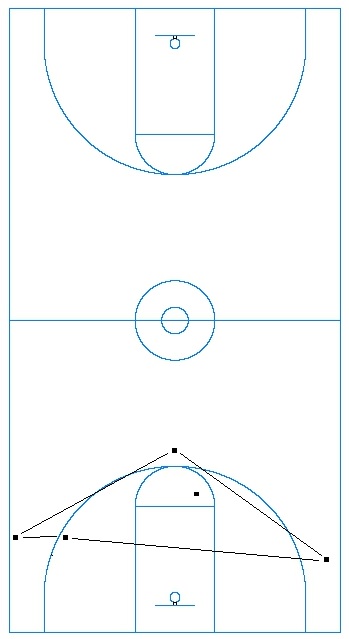}
	\includegraphics[width=0.242\textwidth]{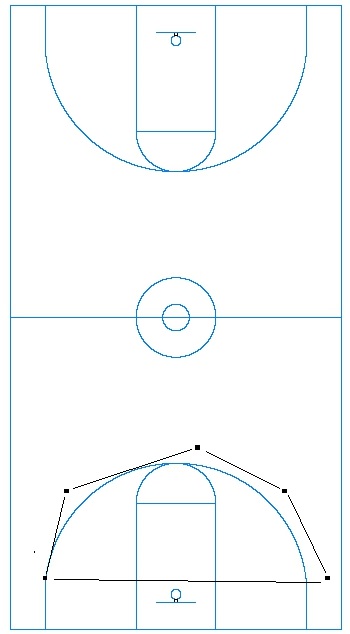}
	\includegraphics[width=0.242\textwidth]{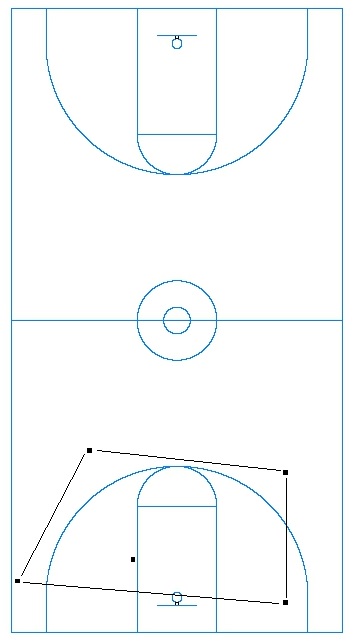}
	\caption{Convex hull for selected snapshots related to the first offensive play of the game}
	\label{CU_A}
\end{figure}

\begin{figure}[!htb]
	\includegraphics[width=0.24\textwidth]{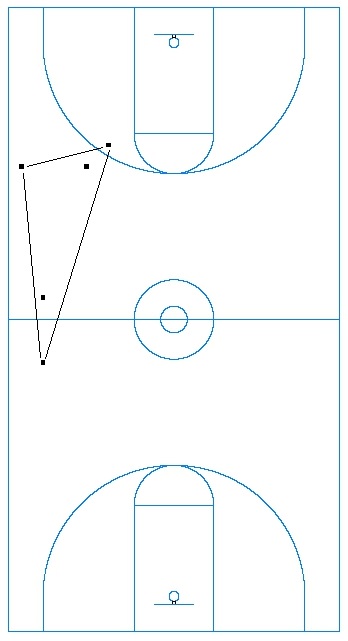}	
	\includegraphics[width=0.24\textwidth]{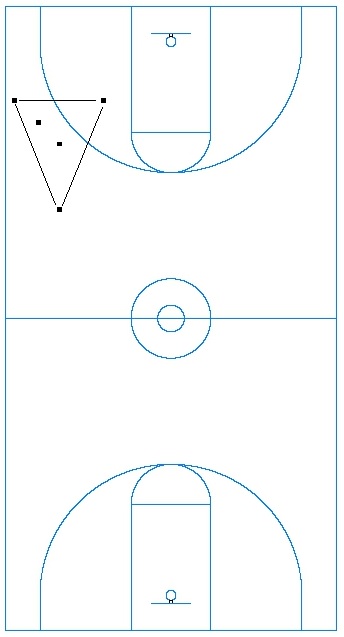}
	\includegraphics[width=0.24\textwidth]{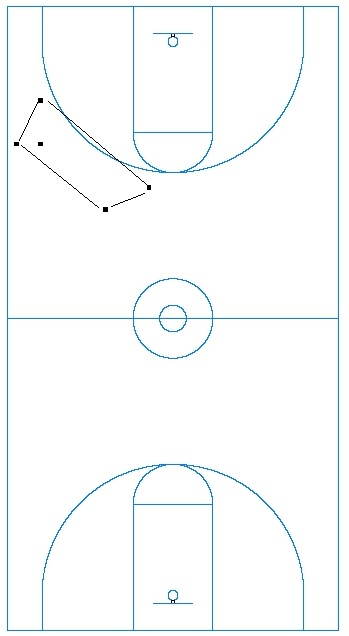}
	\includegraphics[width=0.24\textwidth]{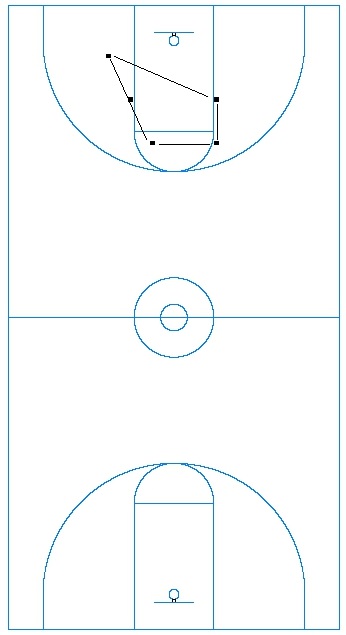}
	\caption{Convex hull for selected snapshots related to the first defensive play of the game}
	\label{CU_D}
\end{figure}

To corroborate the previous evidence, Figures \ref{CU_A} and \ref{CU_D} report the convex hulls for selected snapshots from, respectively, the first offensive play and the first defensive play of the match. Once again, players are more spread around the court in offensive plays.

\begin{figure}[!htb]
	\centering
\includegraphics[scale=0.62]{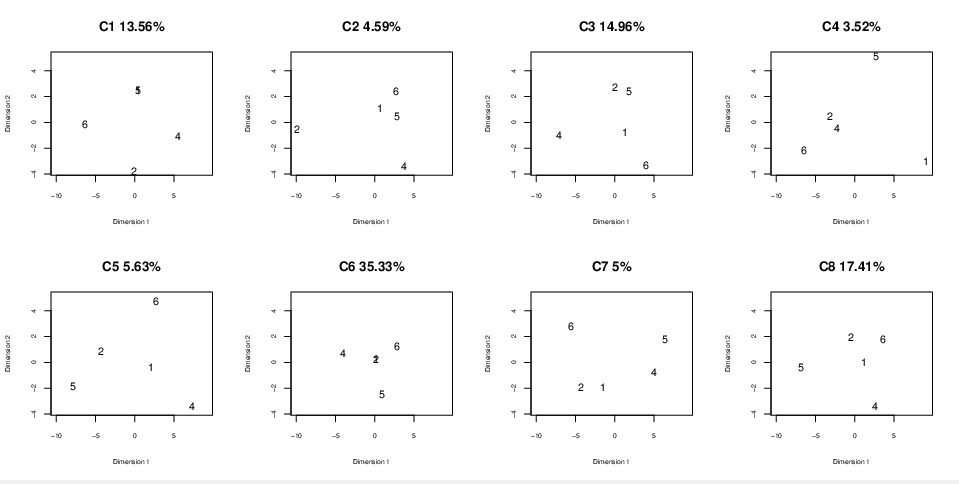}
\caption{Map representing, for each of the 8 clusters (C1: Cluster 1, C2: Cluster 2, ...), the average position in the $x-y$ axes of the five players, using MDS. Percentages report the proportion of instants in the dataset belonging to each cluster.}
	\label{fig:mds}       
\end{figure}

We then perform a cluster analysis, with the aim of characterizing the spatial pattern of the players in the court. We define different game phases, each considering moments being homogeneous in terms of spacings among players. We apply a k-means cluster analysis to group objects. Objects are the time instants and the similarity is expressed in terms of players' distance. We choose k = 8 based on the value of the between deviance (BD) / total deviance (TD) ratio for different number of clusters. First, we characterize each cluster in terms of players' position in the court. For each cluster, a multidimensional scaling (MDS) is used to plot each player in a 2-dimensional space such that the between player average distances are preserved. We find substantial differences among different phases. Results of MDS are presented in Figure \ref{fig:mds} and highlight remarkable differences in the positioning structure of players in the court. 

After having defined whether each moment corresponds to an offensive or a defensive play looking to the average coordinate of the five players in the court, we also found that some clusters represent offensive actions rather than defensive. More precisely, we found that cluster 1 (C1), Cluster 2 (C2), Cluster 3 (C3) and Cluster 4 (C4) mainly correspond to offensive plays (respectively, for the 85.88\%, 85.91\%, 73.93\% and 84.62\% of the time instants in each cluster) and Cluster 6 (C6) strongly corresponds to defensive plays (85.07\%). \textit{Offensive} clusters show larger players' spacings than in the \textit{defensive} cluster.  
A motivation for this behaviour could be that players in defence have the objective to narrow the opponents' spacings in order to limit their play, while the aim of the offensive team is to maintain large distances among team-mates, to increase the propensity to shot with good scoring percentages. Anyhow, these findings go on the same direction of those of the convex hulls.

\section{Conclusions and future developments} \label{concl}

In recent years, coaches, experts and analysts have received benefits from the availability of large amounts of data to use in team sports analysis, which increased the possibility to extract insights from matches in relation to teams' performance. The advent of information technology systems permits to match play-by-play with players' trajectories and to analyse teams' performance with a variety of approaches. Having the trajectories of the players and the play-by-play available, and inspired by the literature based on the data-driven methods as well as by the increasing interest in visualizing data, we analysed the movement and the positioning of players using visual tools and data-mining techniques, with the aim of finding regularities and patterns. 

\vline 

The most promising result relates to convex hulls' analysis. We found that players are more spread around the court in offensive plays rather than in defensive plays. Further analysis should be carried out in order to better understand the logic underpinning this regularity. A potential approach could be to examine the time series of the convex hull areas of both teams together.  This will answer the question whether the defensive team has success in limiting the spacing of the offensive team. An analysis that aims to assess whether and how the two teams pursue their strategies, and how the achievement of their strategy affects their performance, may be of interest for coaches and experts.
 
\vline 

Further research can be carried out with the aim of finding regularities between trajectories and players' (and team) performance by increasing the availability of trajectories data to both team-mates, opponents, and the ball, for multiple games, that sounds essential to better explore the multivariate and complex structure of trajectories in association with teams' performance. Future challenges also aim to experiment the potential of spatial statistics and spatial econometrics techniques applied to trajectory analysis (\cite{brillinger2008modelling}), also in view of the similarities between sport players and economic agents in terms of endogenous and exogenous factors that impact on locational choices, as illustrated in Arbia \cite{arbia2016spatial}.

\section*{Acknowledgements}
Research carried out in collaboration with the Big\&Open Data Innovation Laboratory (BODaI-Lab), University of Brescia (project nr. 03-2016, title Big Data Analytics in Sports, www.bodai.unibs.it/BDSports/), granted by Fondazione Cariplo and Regione Lombardia. Authors would like to thank MYagonism (\texttt{https://www.myagonism.com/}) for having provided the data. A special thank goes to Raffaele Imbrogno ("Foro Italico" University, Roma IV) and Paolo Raineri (MYagonism) for fruitful discussions.

%
%

\begin{thebibliography}{99.}%
%
%

\bibitem{aisch2016curry} Aisch, G., Quealy, K. (2016) \textit{Stephen Curry 3-Point Record in Context: Off the Charts}. New York Times.


\bibitem{arbia2016spatial} Arbia, G. (2016) \textit{Spatial Econometrics: A Broad View}. Foundations and Trends in Econometrics \textbf{8} (3-4), 145-265.

\bibitem{basole2016sports} Basole, R. C., Saupe, D. (2016) \textit{Sports Data Visualization} [Guest editors' introduction]. IEEE Computer Graphics and Applications \textbf{36} (5), 24-26.

\bibitem{bolt2015water} Bolt, M. D. (2015) \textit{Visualizing Water Quality Sampling-Events in Florida}. ISPRS Annals of the Photogrammetry, Remote Sensing and Spatial Information Sciences  \textbf{2} (4), 73.

\bibitem{bradley2007reliability} Bradley, P., O'Donoghue, P., Wooster, B., Tordoff, P. (2007)\textit{The reliability of ProZone MatchViewer: a video-based technical performance analysis system}. International Journal of Performance Analysis in Sport \textbf{7.3} : 117-129.

\bibitem{brillinger2007potential} Brillinger, D. R. (2007) \textit{A potential function approach to the flow of play in soccer}. Journal of Quantitative Analysis in Sports  \textbf{3} (1), 3.

\bibitem{brillinger2008modelling} Brillinger, D. R. (2010) \textit{Modeling spatial trajectories}. Handbook of spatial statistics 463-476.

\bibitem {Carpita et al. (2013)} Carpita, M., Sandri, M., Simonetto, A., Zuccolotto, P. (2013) \textit{Football mining with R}. Data Mining Applications with R.

\bibitem{Carpita et al. (2015)} Carpita, M., Sandri, M., Simonetto, A., Zuccolotto, P. (2015) \textit{Discovering the Drivers of Football Match Outcomes with Data Mining}. Quality Technology \& Quantitative Management \textbf{12.4} 561-577.

\bibitem{catapult2015} Catapult USA, Sports Ltd. (2015) \textit{Wearable Technology for Elite Sports}.\\ URL http://www.catapultsports.com/.

\bibitem{cintia2015harsh} Cintia, P., Giannotti, F., Pappalardo, L., Pedreschi, D., Malvaldi, M. (2015) \textit{The harsh rule of the goals: data-driven performance indicators for football teams}. Data Science and Advanced Analytics (DSAA) 36678. IEEE International Conference pp. 1-10.

\bibitem{Cintia et al. (2015)} Cintia, P., Rinzivillo, S., Pappalardo, L. (2015) \textit{A network-based approach to evaluate the performance of football teams}. Machine Learning and Data Mining for Sports Analytics Workshop, Porto, Portugal.


\bibitem{couceiro2014dynamical} Couceiro, M. S., Clemente, F. M., Martins, F. M., Machado, J. A. T. (2014) \textit{Dynamical stability and predictability of football players: the study of one match}. Entropy \textbf{16} (2), 645-674.

\bibitem{fonseca2012spatial} Fonseca, S., Milho, J., Travassos, B., Araujo, D. (2012) \textit{Spatial dynamics of team sports exposed by Voronoi diagrams}. Human movement science \textbf{31} (6), 1652-1659.

\bibitem{gesmann2013package} Gesmann, M., de Castillo, D. (2013)\textit{ Package `googleVis"}. Interface between R and the Google Chart Tools.

\bibitem{goldsberry2012courtvision} Goldsberry, K. (2012) \textit{Courtvision: New visual and spatial analytics for the nba}. 2012 MIT Sloan Sports Analytics Conference.

\bibitem{goldsberry2013map} Goldsberry, K. (2013)\textit{ Pass Atlas: A Map of where NFL Quarterbacks throw the ball}. Grantland.

\bibitem{heinz2014pratical} Heinz, S. (2014) \textit{Practical application of motion charts in insurance}.

\bibitem{hilpert2011dynamic} Hilpert, M. (2011) \textit{Dynamic visualizations of language change}. International Journal of Corpus Linguistics \textbf{16} (4), 435-461.

\bibitem{impire2015} Impire (2015) Impire AG. URL http://www.bundesliga-datenbank.de/en/products/

\bibitem{losada2016bkviz} Losada, A. G., Theron, R., Benito, A. (2016) \textit{BKViz: A Basketball Visual Analysis Tool}. IEEE Computer Graphics and Applications \textbf{36} (6), 58-68.

\bibitem{metulini2016motion} Metulini, R. (2017) \textit{Spatio-Temporal Movements in Team Sports: A Visualization approach using Motion Charts}. Electronic Journal of Applied Statistical Analysis.

\bibitem{metulini2017cluster} Metulini, R., Manisera, M., Zuccolotto, P. (2017) \textit{Space-Time Analysis of Movements in Basketball using Sensor Data}. in Statistics and Data Science: new challenges, new generations. SIS2017 proceedings (Firenze University Press). e-ISBN: 978-88-6453-521-0.

\bibitem{moura2012quantitative} Moura, F. A., Martins, L. E. B., Anido, R. D. O., De Barros, R. M. L., Cunha, S. A. (2012) \textit{Quantitative analysis of Brazilian football players' organisation on the pitch}. Sports Biomechanics \textbf{11} (1), 85-96.

\bibitem{Passos et al. (2011)} Passos, P., Davids, K., Araujo, D., Paz, N., Minguens, J., Mendes, J. (2011) \textit{Networks as a novel tool for studying team ball sports as complex social systems}. Journal of Science and Medicine in Sport \textbf{14.2} : 170-176. 

\bibitem{araujo2016team} Passos, P., Araujo, D., Volossovitch, A. (2016) \textit{Performance Analysis in Team Sports}. Routledge.


\bibitem{perin2013soccerstories} Perin, C., Vuillemot, R., Fekete, J. D. (2013) \textit{SoccerStories: A kick-off for visual soccer analysis}. IEEE transactions on visualization and computer graphics \textbf{19} (12), 2506-2515.

\bibitem{pileggi2012snapshot} Pileggi, H., Stolper, C. D., Boyle, J. M., Stasko, J. T.: Snapshot (2012) \textit{Visualization to propel ice hockey analytics}. IEEE Transactions on Visualization and Computer Graphics \textbf{18} (12), 2819-2828.

\bibitem{polk2014tennivis} Polk, T., Yang, J., Hu, Y., Zhao, Y. (2014) \textit{Tennivis: Visualization for tennis match analysis}. IEEE transactions on visualization and computer graphics \textbf{20} (12), 2339-2348.

\bibitem{sacha2014feature} Sacha, D., Stein, M., Schreck, T., Keim, D. A., Deussen, O. (2014) \textit{Feature-driven visual analytics of soccer data}. Visual Analytics Science and Technology (VAST), IEEE Conference on (pp. 13-22).

\bibitem{saka2015inequality} Saka, C., Jimichi, M. (2015) \textit{Inequality evidence from accounting data visualisation}.

\bibitem{santori2014liver} Santori, G. (2014) \textit{Application of Interactive Motion Charts for Displaying Liver Transplantation Data in Public Websites}. In Transplantation proceedings \textbf{46} (7), 2283-2286. Elsevier.

\bibitem{santos2012goal} Santos, J. L., Govaerts, S., Verbert, K., Duval, E. (2012) \textit{Goal-oriented visualizations of activity tracking: a case study with engineering students}. In Proceedings of the 2nd international conference on learning analytics and knowledge, pp. 143-152. ACM.

\bibitem{theron2010visual} Theron, R., Casares, L. (2010) \textit{Visual analysis of time-motion in basketball games}. In International Symposium on Smart Graphics. Springer Berlin Heidelberg, pp. 196-207.

\bibitem{tracab2015} Tracab (2015) \textit{Tracab Player Tracking System}. URL http://chyronhego.com/sports-data/player-tracking.


\bibitem{travassos2012spatiotemporal} Travassos, B., Araujo, D., Duarte, R., McGarry, T. (2012) \textit{Spatiotemporal coordination behaviors in futsal (indoor football) are guided by informational game constraints}. Human Movement Science \textbf{31} (4), 932-945.

\bibitem{travassos2013performance} Travassos, B., Davids, K., Araujo, D., Esteves, P. T. (2013) \textit{Performance analysis in team sports: Advances from an Ecological Dynamics approach}. International Journal of Performance Analysis in Sport, \textbf{13.1} : 83-95.

\bibitem{turvey1995toward} Turvey, M. T., Robert E. Shaw. (1995) \textit{Toward an ecological physics and a physical psychology}. The science of the mind: 2001 and beyond, 144-169.

\bibitem{Wasserman and Katherine (1994)} Wasserman, S., Katherine F. (1994) \textit{Social network analysis: Methods and applications}, Vol. 8. Cambridge university press. 

\end{thebibliography}
%

\end{document}